COCCIA LAB

To discover the causes of social, economic and technological change

COCCIALAB WORKING PAPER
2018 – NO. 35/BIS

# SOCIOECONOMIC DRIVING FORCES OF SCIENTIFIC RESEARCH

Mario COCCIA

CNR -- NATIONAL RESEARCH COUNCIL OF ITALY

&

ARIZONA STATE UNIVERSITY

COCCIA*LAB*
at CNR -- NATIONAL RESEARCH COUNCIL OF ITALY
Collegio Carlo Alberto, Via Real Collegio, n. 30-10024, Moncalieri (TO), Italy
*E*-mail: mario.coccia@cnr.it

# Socioeconomic driving forces of scientific research

*Mario Coccia*[1]

CNR -- National Research Council of Italy & Arizona State University

Current Address: Coccia*LAB* at CNR -- National Research Council of Italy
Collegio Carlo Alberto, Via Real Collegio, n. 30, 10024-Moncalieri (Torino), Italy
*E*-mail: mario.coccia@cnr.it

Mario Coccia ORCID: http://orcid.org/0000-0003-1957-6731

Why do nations produce scientific research? This is a fundamental problem in the field of social studies of science. The paper confronts this question here by showing vital determinants of science to explain the sources of social power and wealth creation by nations. Firstly, this study suggests a new general definition of science and scientific research that synthetizes previous concepts and endeavors to extend them: *Science discovers the root causes of phenomena to explain and predict them in a context of adaptation of life to new economic and social bases, whereas scientific research is a systematic process, applying methods of scientific inquiry, to solve consequential problems, to satisfy human wants, to take advantage of important opportunities and/or to cope with environmental threats*. In particular, science and scientific research are driven by an organized social effort that inevitably reflect the concerns and interests of nations to achieve advances and discoveries that are spread to the rest of humankind. This study reveals that scientific research is produced for social and economic interests of nations (power, wealth creation, technological superiority, etc.), rather than philosophical inquiries. A main implication of this study is that the immense growth of science in modern society is not only due to activity of scientists but rather to general social efforts of nations to take advantage of important opportunities and/or to cope with environmental threats, such as war. Empirical evidence endeavors to support the sources of scientific research for nations, described here. Finally, relationships between R&D investment and productivity, and research policy implications are discussed.

[1] I gratefully acknowledge financial support from the CNR - National Research Council of Italy for my visiting at Arizona State University where this research started in 2016 (Grant CNR - NEH Memorandum n. 0072373-2014 and n. 0003005-2016).

## What is science and scientific research?

The purpose of this study is to criticize the motivations of nations to do scientific research to explain and generalize properties over time and space. Before discussing these topics, the study here clarifies the concept of science and scientific research.

The term science has different meanings. Science is an accumulation of knowledge and includes basic and applied fields of research (Coccia and Wang, 2016; Godin, 2001). The Scottish philosopher Rae (1834, p. 254) states that: "the aim of science may be said to be, to ascertain the manner in which things actually exist". A different definition of science was given by Crowther (1955): "Science is a system of behavior by which man acquires mastery of his environment". Volta (1792)[2] considered science in an experimental perspective that has its greatest and most rewarding moments in practical activity. As a matter of fact, science for Volta (1792) is invention and it is driven by scientists' aptitude and/or passion for the construction of new devices and artefacts. Bernal (1939, p. 6) considered science "the means of obtaining practical mastery over nature through understanding it". Instead, Dampier (1953) claimed that science is: "Ordered knowledge of natural phenomena and the rational study of the relations between the concepts in which those phenomena are expressed". Russell (1952) provided a broader definition of science: "Science, as its name implies, is primarily knowledge; by convention it is knowledge of a certain kind, namely, which seeks general laws connecting a number of particular facts. Gradually, however, the aspect of science as knowledge is being thrust into the background by the aspect of science as the power to manipulate nature". According to Freedman (1960) the definition by Russell (1952) is the more satisfactory, while Dampier's definition relates only to scientific knowledge, and does not take into account either the application of such knowledge, or the power to apply it towards control and change of man's environment. However, Russell (1952) describes science as static, whereas it is a dynamic process.

[2] Alessandro Volta (1745-1827) Italian physicist, known for his pioneering studies in electricity. He also invented the electric battery in 1800.

Kuhn (1962) states that:

> Science is a constellation of facts, theories, and methods… Hence scientific development is the fragmentary process through which these elements have been added, singularly or in groups, to the ever growing depository that constitutes technical and scientific knowledge.

Lakatos (1968, p. 168, original Italics and emphasis) argues that:

> science . . . can be regarded as a huge research program . . . .progressive and degenerating problem-shifts in series of successive theories. But in history of science we find a continuity which connects such series. . . . The programme consists of methodological rules: some tell us what paths of research to avoid (*negative heuristic*), and others what paths to pursue (*positive heuristic*) - By 'path of research' I mean an objective concept describing something in the Platonic 'third world' of ideas: a series of successive theories, each one 'eliminating' its predecessors (in footnote 57) - . . . . What I have primarily in mind is not science as a whole, but rather particular research-programmes, such as the one known as 'Cartesian metaphysics. . . . a 'metaphysical' research-programme to look behind all phenomena (and theories) for explanations based on clockwork mechanisms (positive heuristic). . . A research-programme is successful if in the process it leads to a progressive problem-shift; unsuccessful if it leads to a degenerating problem-shift . . . . Newton's gravitational theory was possibly the most successful research-programme ever (p. 169). . . . The reconstruction of scientific progress as proliferation of rival research-programmes and progressive and degenerative problem-shifts gives a picture of the scientific enterprise which is in many ways different from the picture provided by its reconstruction as a succession of bold theories and their dramatic overthrows (p. 182).

Considering these different perspectives, Freedman (1960, p. 3) suggests the following definition of science:

> Science is a form of human activity through pursuit of which mankind acquires an increasingly fuller and more accurate knowledge and understanding of nature, past, present and future, and an increasing capacity to adapt itself to and to change its environment and to modify its own characteristics.

This study argues that:

| |
|---|
| *Science discovers the root causes of phenomena to explain and predict them in a context of adaptation of life to new economic and social bases.* |

Table 1 synthetizes some definitions of science and scientific research given by scholars.

Table 1. Scholars and suggested definition of science

| *Authors (year)* | *Suggested definition of science and scientific research* |
|---|---|
| Volta (1792) | Science has its greatest and most rewarding moments in practical activity and is driven by scientists' aptitude for the construction of new devices and artefacts |
| Rae (1834) | The aim of science is to ascertain the manner in which things actually exist |
| Bernal (1939) | Science is the means of obtaining practical mastery over nature through understanding it |
| Crowther (1955) | Science is a system of behavior by which man acquires mastery of his environment |
| Dampier (1953) | Ordered knowledge of natural phenomena and the rational study of the relations between the concepts in which those phenomena are expressed |
| Russell (1952) | Science is primarily knowledge; by convention it is knowledge of a certain kind, namely, which seeks general laws connecting a number of particular facts. …the aspect of science as knowledge is being thrust into the background by the aspect of science as the power to manipulate nature |
| Freedman (1960) | Science is a form of human activity through pursuit of which mankind acquires an increasingly fuller and more accurate knowledge and understanding of nature, past, present and future, and an increasing capacity to adapt itself to and to change its environment and to modify its own characteristics. |
| Kuhn (1962) | Science is a constellation of facts, theories, and methods… Hence scientific development is the fragmentary process through which these elements have been added, singularly or in groups, to the ever growing depository that constitutes technical and scientific knowledge. |
| Lakatos (1968) | Science . . . can be regarded as a huge research program . . . .progressive and degenerating problem-shifts in series of successive theories. But in history of science we find a continuity which connects such series. . . . |
| Coccia (2018, this paper) | Science discovers the root causes of phenomena to explain and predict them in a context of adaptation of life to new economic and social bases. |

These different views of science show that the concept of science is elusive and a definition of science is a hard task because of the nature of science itself. In this background of social studies of science, it is possible to clarify the concepts of research and scientific research. Generally speaking, research is continued search for knowledge and understanding in society. Instead, scientific research is a continued search for advancing scientific knowledge, applying methods of inquiry.

This study considers scientific research as: scientific research is a systematic process, applying methods of scientific inquiry, to solve consequential problems, to satisfy human wants, to take advantage of important opportunities and/or to cope with environmental threats. In addition, scientific research, as a systematic process, is driven by an organized social effort of nations to make science advances and discoveries known to the rest of humankind.

The dual elements of the scientific nature of a research are: determination of problems and utilization of the methods of inquiry (they are organized and systematic scientific thinking used by scholars for controlled investigations and experiments to logically and efficiently solve theoretical and practical problems, and generate discoveries and/or science advances, see Coccia, 2018g).

In particular, scientific research can be carried out with following general methods of inquiry (Coccia, 2018g):

- *Inductive approach* starts from the experimental observation of phenomena and traces back the laws that regulate them by means of experiments, analogies, and hypotheses;
- *Deductive approach* starts from theory and general ideas in order to predict new laws and explain new phenomena.

The process of scientific research can be described with the theoretical framework of the Gestalt psychology given by (see Basalla, 1988, p. 23; cf., Usher, 1954): 1) Perception of the problem: an incomplete pattern in need of resolution is recognized; 2) Setting stage: data related to the problem is assembled; 3) Act of insight: a mental act finds a solution to the problem; 4) Critical revision: overall exploration and revision of the problem and improvements by means of new acts of insight[3].

[3] For studies about the role of science, technology, sources of innovation and knowledge in society, see also, Calabrese et al., 2002, 2005; Calcaltelli et al., 2003; Cavallo et al., 2014, 2014a, 2015; Chagpar and Coccia, 2012; Coccia, 2001, 2002, 2003, 2004, 2004a, 2005, 2005a, 2005b, 2005c, 2005d, 2005e, 2005f, 2005g, 2005h, 2006, 2006a, 2008, 2008a, 2008b, 2009, 2009a, 2009b, 2009c, 2009d, 2010, 2010a, 2010b, 2010c, 2010d, 2010e, 2010f, 2011, 2012, 2012a, 2012b, 2012c, 2012d, 2013, 2013a, 2014, 2014a, 2014b, 2014c, 2014d, 2014e, 2014f, 2015, 2015a, 2015b, 2015c, 2016, 2016a, 2016b, 2017, 2017a, 2017b, 2017c, 2017d, 2017e, 2017f, 2017g, 2017h, 2017i, 2017l, 2018, 2018a, 2018b, 2018c, 2018d, 2018e, 2018f; Coccia and Bozeman, 2016; Coccia and Cadario, 2014; Coccia and Finardi, 2012; Coccia et al., 2010, 2012, 2015; Coccia and Rolfo, 2002, 2007, 2008, 2009, 2010, 2013; Coccia and Wang, 2015, 2016; Rolfo and Coccia, 2005; Benati and Coccia, 2017, 2018; Coccia and Benati, 2018; 2018a.

Although several contributions in social studies of science, the problem of why nations sustain science and scientific research is hardly clarified. In particular, which complex factors drive nations to support science and scientific research are basic to explain human development in society (Coccia and Bellitto, 2018). In light of the continuing importance of these topics in the social studies of science, this paper seeks to explain critical factors supporting nations to produce science and scientific research in society.

**Why do nations produce scientific research in society?**

Scientific research reflects the social climate in which it is carried out. Most of the significant discoveries are a systematic, generally organized process of scientific research that reflects the outward-looking tendencies in society. Bernal (1939) analyzed the social function of science considering its practical activities as the basis of progress. Bernal (1939) also argued that science is produced for social and economic interests of nations rather than a philosophical inquiry. A main implication is that the immense growth of science in modern society is not only due to activity of scientists but rather to general social efforts of nations to take advantage of important opportunities and/or to cope with environmental threats, such as war. In general, scientific research has been less a matter of individual enterprise and more an organized social effort (Coccia and Wang, 2016). Social climate of nations affects the development of scientific research, the understanding and appreciation of scientific discoveries in society. Scientists inevitably reflect the concerns and interests of their home society. Figure 1 shows some factors affecting the production of scientific research by nations and next sections endeavor to explain these factors.

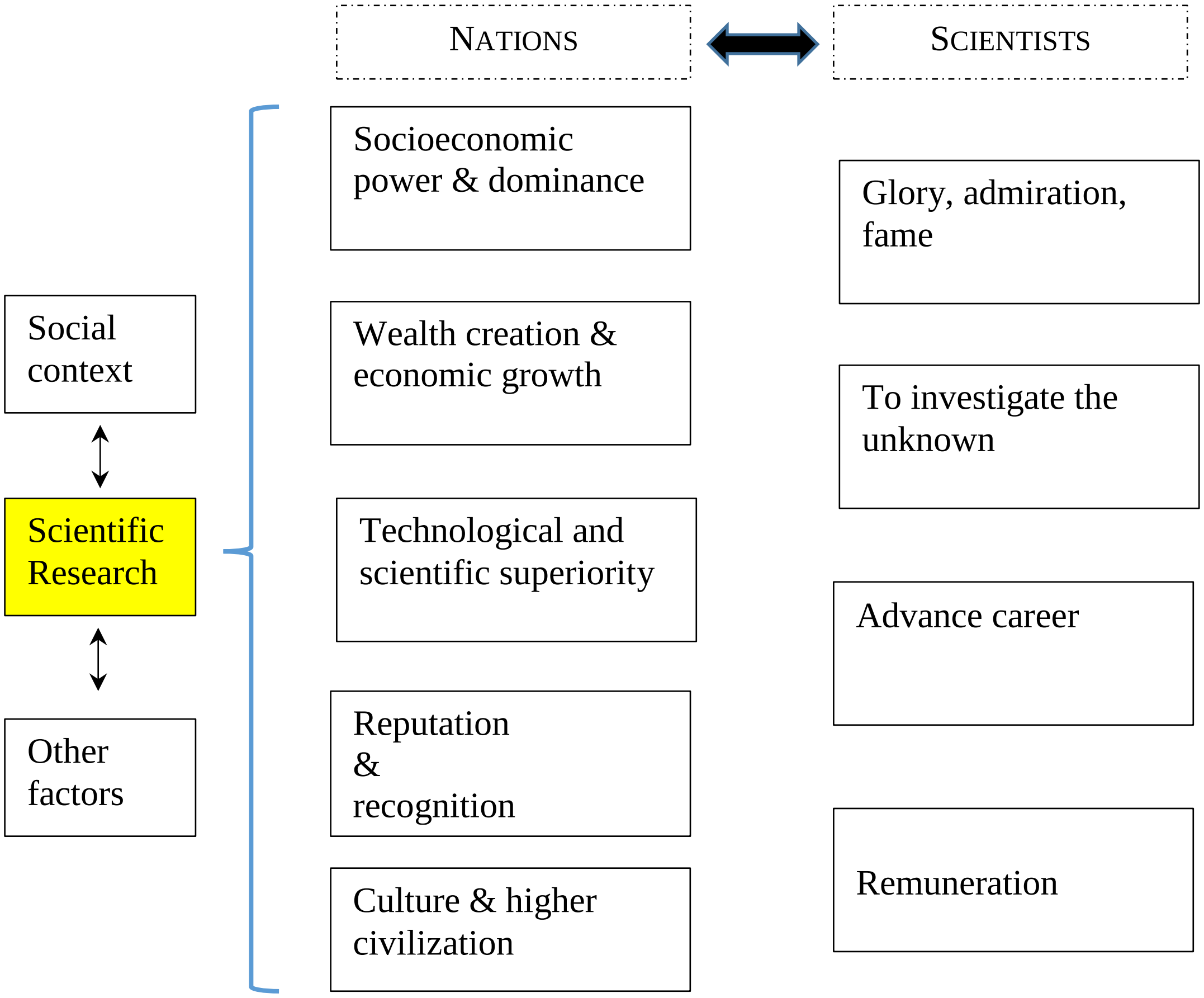

**Figure 1.** Factors associated with the production of scientific research by nations and scientists

### *1.1 Scientific research as a source of socioeconomic power*

A nation can perform scientific research to support a socioeconomic power directed to take advantage of important opportunities and/or to cope with consequential environmental threats, such as war. Socioeconomic power of a nation is based on a process of influence on other subjects towards the accomplishments of some goals (e.g., mutual trade), in some cases associated with (formal and/or informal) dominance and control of geoeconomic areas. Scientific research can generate achievements that are also important in the presence of socioeconomic shocks, such as warfare (cf., Ruttan, 2006; Constant, 2000; Mowery, 2010). The investigation of war economy and mainly of war consequences can help to understand the reasons why nations perform scientific research. A main purpose of societies in war is to take advantage of opportunities to have fruitful socioeconomic consequences and gain dominance and control on other areas. In the Ancient period, the victory in war was due to the strength and prowess of population, whereas the modern warfare depends more and more on scientific, technical and engineering knowledge of nations (Coccia, 2015; 2017). Current international conflicts are won in research labs with high-tech weapons and cyberpower (cf., Kramer *et al.*, 2009). The pioneering studies by Neurath (1919) showed the stimulating effect of war on technical and scientific progress of countries. Recently, some social scientists have paid more attention to the effects of scientific research on technology during war and post war period (cf., Coccia, 2015, 207, 2018; Ruttan, 2006; Mowery, 2010). War can support not only scientific research but also other types of novelties, such as innovative laws and regulations. Moreover, social scientists have a theoretical reluctance to differentiate between types of warfare. The tendency is to treat war as a generic phenomenon with equivalent socioeconomic impact, whereas some wars are more important than others in terms of impetus for nations to produce scientific research, discoveries and new technology. In particular, there is a distinctiveness of world war, which generates major socioeconomic consequences and many science advances by countries to gain dominance and global leadership (Stein and Russett, 1980, p. 401; Coccia, 2015).

Nations support scientific research to have a high *economic potential* based on a scientific and technological superiority both in peacetime and in warfare period (cf., Mendershausen, 1943, p. 8; Smith, 1985). Recent studies by Ruttan (2006) analyze the relation among war, science, innovation and economic growth of countries. Ruttan (2006, p. 184*ff*) argues that without a *threat* of a major war, it is difficult that the U.S. political system mobilizes huge human and economic resources to support the development of major and strategic discoveries that subsequently can be translated in commercial innovations for the progress in society. In short, the fruitful factors at the origin of vital discoveries and science advances thrive in the presence of international conflicts and crises, driven by common institutional, entrepreneurial and scientific energies, to cope with consequential environmental threats. *Innovative spirit* guide scientific research of countries in the presence of war, based on two critical drivers: *demand factors* spur a huge demand shock because of a massive increase in deficit spending with expansionary policy (cf., Field, 2008); *supply factors*: learning by doing in military production, spin-off and spillover from military R&D, etc. Wright (1997, p. 1565) examines the “American technological leadership” and shows that critical manufacturing sectors for U.S. economy[4] have taken advantages from fruitful demand- and supply-side effects of wars (cf. also, Goldfarb, 2005). The mobilization for wars increases R&D investments to produce sciences advances associated with military technologies that are transferred to civilian applications in the long term to support a higher economic potential and economic growth (Goldstein, 2003; Stein and Russett, 1980, p. 412). In particular, a strong economic and scientific potential has a vital role to win wars for the distribution of power within the international system (Modelski, 1972; cf., Levy, 2011). Modelski (1972, p. 48) asserts that the “war causes the Great Powers”, which affect the political and economic system worldwide (e.g., Roman Empire over 200BC ~ 400AD, Britain Empire in the 1710-1850 period, the USA from 1940s onwards, etc.; cf., Stein and Russett, 1980). In fact, Ferguson (2010) claims that the United States has a global leadership because of a stronger military, political, scientific, technological and economic potential worldwide recognized.

Instead, Coccia (2015, 2017) suggests that sources of science and technology are, *de facto*, associated with the goal of global leadership of purposeful systems (e.g., nations) in the presence of effective and/or potential environmental threats, rather than warfare *per se*. In short, the source of major science advances seems to be driven by solution of relevant and strategic problems -in the presence of consequential environmental threats to national security-, in order to achieve/sustain/defend the position of global leadership by nations.

Table 2 shows that nations, such as the USA having higher investments in R&D, generate higher innovative outputs and GDP per capita than other nations: these factors are proxies of socioeconomic power. Moreover, Coccia (2015, 2017) shows that U.S. Department of Defense had about 700 foreign installations in 2000s in more than 60 countries worldwide (U.S. DoD, 2003, 2012). The high presence of U.S. military installations confirms the U.S. global leadership, achieved winning World War II, associated with a high economic, scientific and technological potential worldwide recognized (Coccia, 2015). As a matter of fact, nations invest in scientific research to support new technology to be more efficient in the presence of effective and/or potential international conflicts, environment threats and across markets; for instance, military and political tensions between U.S. and Soviet Union in the 1960s, during the period of Cold war, have supported a high investment in scientific research that has generated many discoveries and new technology in order to prove scientific and technological superiority worldwide, and military strength in space (cf., Kira and Mowery, 2007; Ruttan, 2006).

[4] For instance: aircraft, electrical machinery, non-electrical machinery, chemicals and allied products, and motor vehicles.

*Table 2.* R&D investments and innovative output of leading nations to support socioeconomic power worldwide

| Countries | Average Military expenditure by country as percentage of gross domestic product 1992-2013 * | Average Research and Development expenditure (% of GDP) 1996-2005 ϕ | Average Patent applications, residents per million People 1985-2005 ϕ | Average GDP per capita, PPP (constant 2005 international $) 1989-2006 ϕ |
|---|---|---|---|---|
| United States | 3.90 | 2.66 | 447.20 | 36,318.11 |
| Russia/USSR | 3.87 | 1.09 | 145.84 | 9828.36 |
| France | 2.64 | 2.18 | 224.04 | 27,439.67 |
| UK | 2.60 | 1.82 | 334.51 | 26,565.94 |
| China P. R. | 1.99 | 0.92 | 18.00 | 2,398.01 |

*Note*: * SIPRI Military Expenditure Database (2012); ϕ World Bank (2008).

## 1.2 *Scientific research as a source of economic growth and competitive advantage of nations*

Bacon (1629) [5] believed that science had the power to improve the society's economy and standard of living. In his work *New Atlantis* (Bacon, 1629), he saw science, technology, politics, industry, and religion as deeply intertwined. Stephan (1996, p. 1199) argues that science is one of the sources of economic growth. In particular, science supports technological innovations and has interrelationships with economic growth and other socioeconomic forces (Coccia, 2017, 2018).

[5] Bacon is known as the father of the English empiricist philosophy, a tradition that includes Locke, Hume, J.S. Mill, Russel.

The endogenous growth theory is one of the most prominent developments in the field of economic theory (Nelson and Romer, 1996). Romer (1994) and Lucas (1988) argue that economic growth depends on – i.e., it is endogenous to – investments in scientific research and education. The endogenous growth theory is influencing modern economic policies of both industrialized and emerging countries, since investments in higher education, as well as in R&D of firms and public research organizations are vital elements for the increase of new technology, productivity and economic growth within national innovation systems (Coccia, 2004, 2005h, 2011, 2013, 2013a, 2016; Coccia et al., 2015; Coccia and Cadario, 2014; Coccia and Rolfo, 2002, 2009, 2010, 2013; Larédo and Mustar, 2004). However, Bernal (1939), writing between the two World Wars, was not optimistic about science. Barnal's work explicitly recognizes the lack of direct links between social and scientific progress. In fact, science advances, associated with technological progress, can also generate negative effects, such as a higher pollution and incidence of cancer in society (Coccia and Bellitto, 2018). Coccia (2015b) seems to reveal a main interrelationship between high scientific, technological and economic performance (indicators of human progress) and high diffusion of some cancers between countries, controlling screening technology (e.g., computed tomography).

### *1.3 Scientific research as a source of new technology*

One of the reasons to invest in R&D is to generate new technology that, in turn, supports competitive advantage of firms and nations (Porter, 1985; 1990). This argument can be explained with the linear model by Bush (1945):

*basic physics→large scale development→applications→military and civil innovations* [1]

Linear model of R&D [1] considers a stepwise progression from basic science, discoveries through applied research to technological development in firms and research labs, leading to a cluster of new products for wellbeing in society. Rothwell (1994, p. 40, original emphasis) argues that the underlying reason that leads nations to invest in scientific research is that "more R&D in 'equalled' more innovation out". The model [1] is improved over time with a more general process of coupling between science, technology and market, as well as systems integration and networking within and between public and private R&D laboratories directed to produce scientific research and new technology, which are beneficial for society and its wellbeing. Bush (1945) also suggests that basic science should be publicly funded and left to itself in order to produce advances in applied science and technology. This perspective was influential on the post-war research policy in a period of accelerated economic growth (Bush, (1945). Callon (1994) argues that public subsidy to support emerging research fields is needed, though results can be uncertain and/or achieved only in the long run, such as in gravitational astronomy that studies the sources of the universe. De Solla Price (1965) recognizes the interaction between science and technology and uses the metaphor of two dancing partners who are independent but move together (cf., de Solla Price, 1963; Kitcher, 2001). Finally, Gibbons and Johnston (1974) argue that scientific research of nations generates value that can be applied to solve specific problems, translating the results of scientific research in industrial environment for increasing employment and wealth of nations.

### *1.4 Scientific research to increase reputation and recognition within and between scientific communities and nations*

Stephan and Levin (1992) and Stephan and Everhart (1998) argue that scientists in their social context are interested in three types of rewards :

1) the game, the satisfaction derived from solving a problem and investigating the unknown. Hull (1988, p. 305) describes scientists as being innately curious to investigate the unknown to achieve glory, fame and recognition. However, the activity of scientists, research teams, universities and research labs reflect an organized social effort of nations in specific historical periods (Stephan, 1996).

2) the glory and fame: the prestige that accompanies priority by scientists and nations in discovery. Merton (1957, 1968, 1972) argues that the goal of scientists and nations is also to establish priority of discovery by being first to communicate an advance in science worldwide. Publication is a lesser form of recognition, but a necessary step in establishing priority knowledge and that the rewards to priority are the recognition awarded by the scientific community and other nations for being first (Stephan, 1996). Dasgupta and Maskin (1987) argue that there is no value added when the same discovery is made a second, third, or fourth time. To put sharply, the winning research unit is the sole contributor to social surplus. Zuckerman (1992) estimates that, in the early 1990s, around 3,000 scientific prizes were available in North America alone to support recognition of scholars and research labs. A defining characteristic of winner-take-all contests is inequality in the allocation of rewards. In fact, scientific research generates extreme inequality with regard to scientific productivity and awarding priority. This feature also generates the high productivity of some researchers and universities (e.g., MIT, Harvard University, Yale University, etc.) based on cumulative learning processes, called Matthew effect in science (Merton, 1957). This effect shows that researchers/research labs/universities who accomplish prominent results at the beginning of their history have an initial advantage over others and increased chances of obtaining further financial support as well as of accomplishing further discoveries.

3) the monetary rewards. Financial remuneration is another component of the reward structure of science. Compensation in science is generally composed of two parts: one portion is paid regardless of the individual's success in races, the other is priority-based and reflects the value of the winner's contribution to science. While this clearly oversimplifies the compensation structure, the role played by counts of publications and citations in determining raises and promotions at universities is evident from the work by Diamond (1986). Moreover, discoveries and science advances generate patents that are a main source of money that leads to new technology supporting employment and competitiveness of nations worldwide (Jaffe and Trajtenberg, 2002).

### *1.5 Scientific research as a source of profit and socioeconomic problems of marketization in science*

The connection between science and industry supports economic growth and progress (Coccia, 2012b). Rosenberg (1974) argues that science produces advances in scientific knowledge that can reduce the cost of solving complex technological problems and the cost of producing new technology. Mansfield (1995) shows that scientific research has a main impact on innovative products and processes in industry (cf., Jaffe and Trajtenberg, 2002). He also shows that some high-tech sectors have fruitful interactions between technology and basic sciences. Moreover, many nations support a growing commercialization of scientific research and technology transfer to support profit of firms (Slaughter and Leslie 1997; Coccia, 2004, 2009b). The commercialization of scientific research for maximization of profits by firms is driven by efficient R&D labs (Coccia, 2016a). For instance, leading firms in biopharmaceutical sectors invest in Research and Development (R&D) a high level of economic and human resources to support new knowledge and drug discovery to maximize the profit with new compounds (Coccia, 2014f, 2015c, 2018f), such as:

- AstraZeneca (UK-Sweden) invested about US$ 4 billion in 2012
- Roche (Switzerland) about US$ 10.6 billion US
- Boehringer Ingelheim (Germany) about $ 4.3 billion euro of R&D investments

In current competitive markets, public research labs have also a market orientation with many characteristics of business firm (cf., Coccia, 2012e). However, this phenomenon has been criticized because "the embracement of the market is compromising scientific norms and commercialization (or commodification, or marketization) is in profound conflict with the function and main mission of research units and universities" (Musselin 2007; cf. also Greenfeld, 2001), that is, knowledge creation through research and dissemination through publication and education (Schuetze 2007; Slaughter and Leslie 1997). Washburn (2005) offers a highly critical assessment of close science and industry ties for profit maximization, showing "the great and dangerous influences that money and corporate ties impose." The "massification" of scientific research, associated with business and commercial interests, is influencing science in an "unsavory manner." Nelson (2005) states that "there are real dangers that unless [marketization of the scientific research] is halted soon, important portions of future scientific knowledge will be private property and fall outside the public domain [and] that could be bad news for future progress of science and for technological progress." The risk of this tendency, according to Laudel (2006), is that basic research and knowledge might suffer. Certain lines of basic research, whose success is difficult to predict, might become "endangered species" (Laudel 2006). Such forebodings are relevant to modern, knowledge-driven economies in their support R&D management to foster academic institutions and labs that play a driving role as "engines of growth," based on their intangible capital, brainpower. In this context, Rosenberg and Birdzell (1990) argue that science pushes the frontiers of knowledge creating economic resources for firms and nations. However, science advances can also increase the economic gap between countries that apply a Western-style of production and others not applying it.

### *1.6 Public and private scientific research for supporting productivity of nations*

Scientific research and innovation take up considerable economic and human resources that contribute to the accumulation of intangible capital of countries for long-term economic growth (Lucas, 1988; Romer, 1994; Porter, 1985, 1990). R&D investments are a main indicator of the level of science and scientific research of nations (Coccia, 2008a, 2012b). Several studies confirm the positive influence of Research & Development (R&D) expenditure on the growth of productivity of nations (Mairesse and Sassenou, 1991; Amendola et al., 1993; Hall and Mairesse, 1995; OECD, 2003). Many studies aim at understanding whether public investment in R&D is a complement or substitute for R&D private investment (Blank and Stigler, 1957; Kealey, 1996; Coccia, 2010b, 2010e) but, despite the vast scientific literature, results are rather ambiguous. Some studies show that public financing has spillover effects on private investments in R&D (Adams, 1990; Jaffe, 1989; Toole, 1999). In particular, Grossman and Helpman (1991) show that spillovers from R&D are an important source of growth. Other studies show how public and private R&D investments influence the productivity of countries (Levy and Terleckyj, 1983). Lichtenberg and Siegel (1991) and Hall and Mairesse (1995) provide indications of the correlation between R&D investment and productivity. Amendola et al. (1993) present well-documented evidence that R&D investment has noticeable effects on the growth of both productivity and competitiveness of nations. According to Brécard et al. (2006), R&D produces effects on aggregate productivity gains. Griffith et al. (2004) claim that R&D has a direct effect on the growth of Total Factor Productivity (TFP) in a panel of sectors for 12 OECD countries. Aghion and Howitt (1998) claim that R&D investment causes productivity growth, which in turn supports the Gross Domestic Product (GDP). Zachariadis (2004) uses aggregate data from manufacturing sector for a group of OECD countries in 1971-1995 and he finds that R&D intensity has a positive impact on growth rates of both productivity and GDP. Zachariadis (2004), Guellec and van Pottelsberghe de la Potterie (2004) also show the positive relationship between TFP and R&D investments. About the relation between public and private R&D investments, Wallesten (1999) gives evidence for a crowding-out effect, whereas Robson (1993) claims that there is one-to-one complementarity. Blank and Stigler (1957) use a sample of firms to show that there is a substitution effect, but by changing the sample they find a complementarity effect. David et al. (2000) argue that 1/3 of the case studies at firm, sector, and aggregate levels show a substitution effect of public research expenditure for private investments.

A complete analysis of the substitution or crowding out effect of R&D expenditure is necessarily related to the understanding of the decision mechanisms used by public bodies (governments and departments) and private subjects (e.g., firms). Coccia (2010b, 2010e) shows that at the aggregate level, the complementarity between public and private R&D investment but it is important for the government to support a level of public R&D expenditure, as part of the total GDP, lower than that of business R&D investment in order to drive productivity and economic growth in the long run. Therefore, in order to produce positive effects at national level, public R&D expenditure should be lower than the firms' expenditure to avoid crowding out effects. Moreover, high public R&D financing can be counterproductive and increase public deficit, with negative repercussions on interest rates and country's future economic performances (cf., Coccia, 2017i). Steil et al. (2002) claim that in the USA, Japan, Germany, France, and the UK, the interventionist role of the government in the economic field has reduced in favor of that of the market forces, which have become more important in the allocation of resources within the research sector, even though several governments have not yet solved problems regarding under-investments in basic research, which is a public good (Arrow, 1962). In 2002, the European Union induced European countries, in line with international trends, towards an increase in R&D investments: the goal was 3% of the GDP, 56% of which should be financed by the private sector, in order to achieve the innovation intensity and growth levels of the USA by 2010 (European Commission, 2003; 2004; 2005; Room, 2005). This result could have been achieved if governments had implemented a range of incentives to private firms to stimulate their industrial R&D investments. In particular, governments should encourage industrial research labs of firms to recruit scientists and engineers from universities and public labs, so that the economic system has more industrial scientists and fewer academic scientists. In 2018, the ambitious target of 3% of R&D/GDP within EU countries is fail due to economic turmoil in 2000s and socioeconomic problems of high public debt within many countries (Coccia, 2017i).

Coccia (2010b, 2010e) confirms high economic performances in countries with low public financing to R&D associated with high investments in research by private enterprises (e.g., in the UK, the USA, Germany, etc.). Private firms are capable of investing in a much better way than the Government, the politicians, and the bureaucrats do for increasing employment, economic growth and wealth of nations (Coccia, 2010e). Figures 2-4 show low economic performances in countries (for example Italy) whose public expenditure in R&D is higher than private expenditure. In brief, the public policy of stimulating private investments in research rather than public R&D investments, it increases labor productivity per hour worked and long-term economic growth. The effects of these research policies are amplified when combined with economic stability, effective regulations, liberalizations, and competition policies.

Coccia (2009a) also shows that the range of gross domestic expenditure on R&D expressed as percentage of GDP (GERD) between 2.3 per cent and 2.6 per cent maximizes the long-run impact on productivity growth and it is the key to sustained productivity and technology improvements that are becoming more and more necessary to modern economic growth. Moreover, Coccia (2018f), based on OECD data, reveals that (very) high rates of R&D intensity and tax on corporate profits do not maximize the labor productivity of nations. In particular, the models suggest that the R&D intensity equal to about 2.5% and tax on corporate profits equal to 3.1% of the GDP seem to maximize the labor productivity of OECD countries (Figg. 5 and 6).

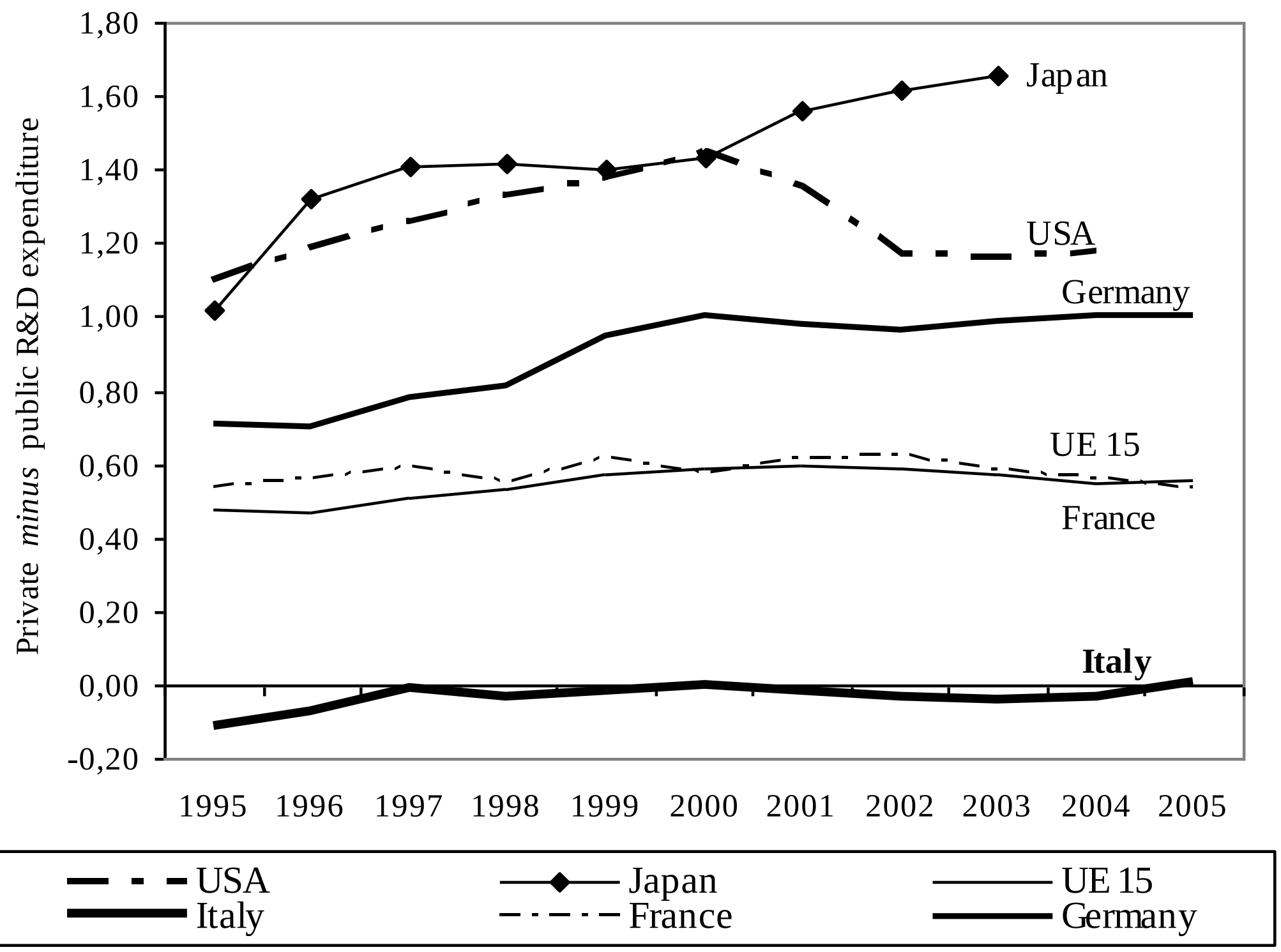

Figure 2 – Private minus public R&D expenditure over time per country. *Source*: Coccia, 2010b; 2010e

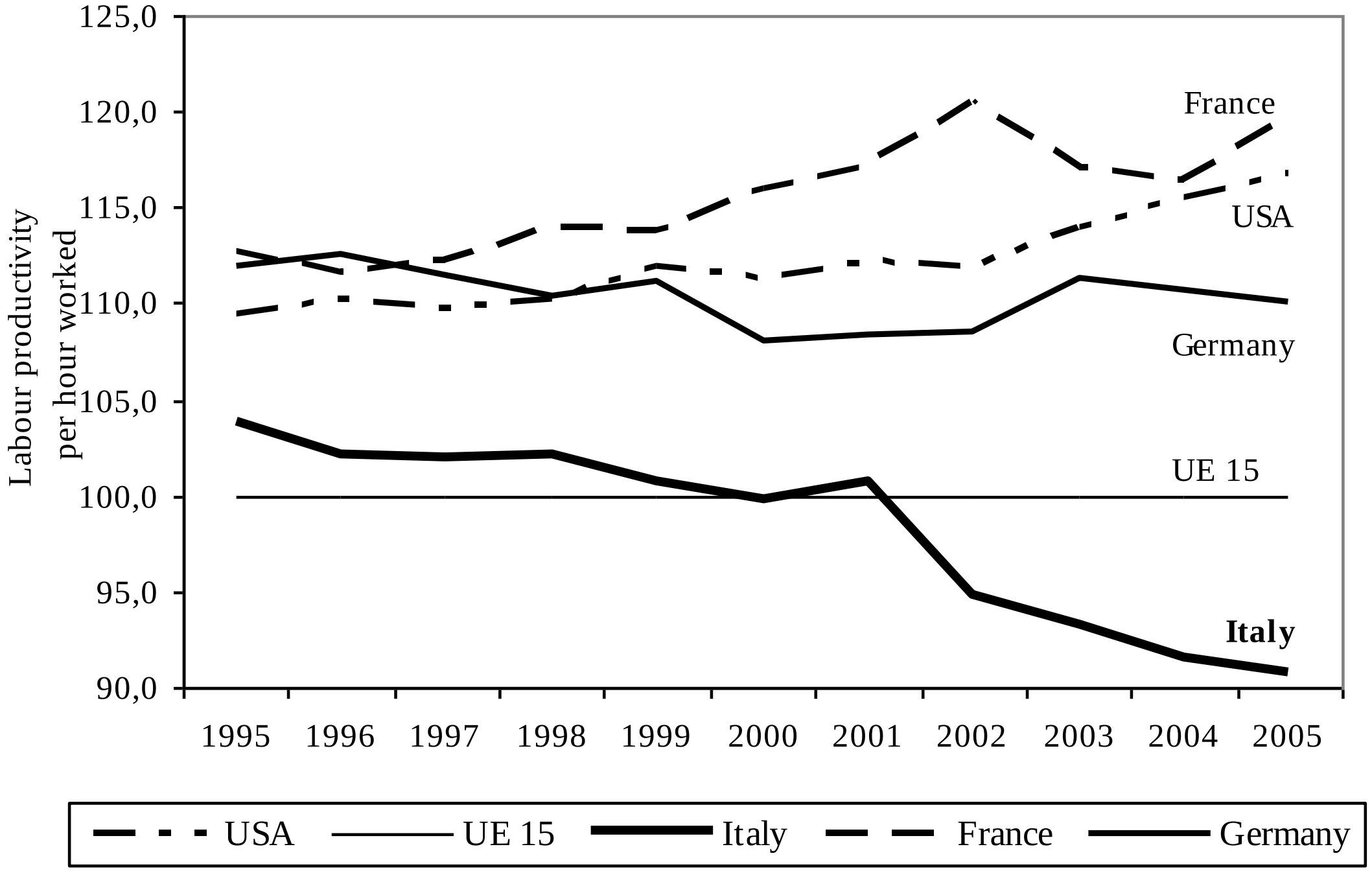

Figure 3 – Labor productivity per hour worked over time per country. *Source*: Coccia, 2010b; 2010e

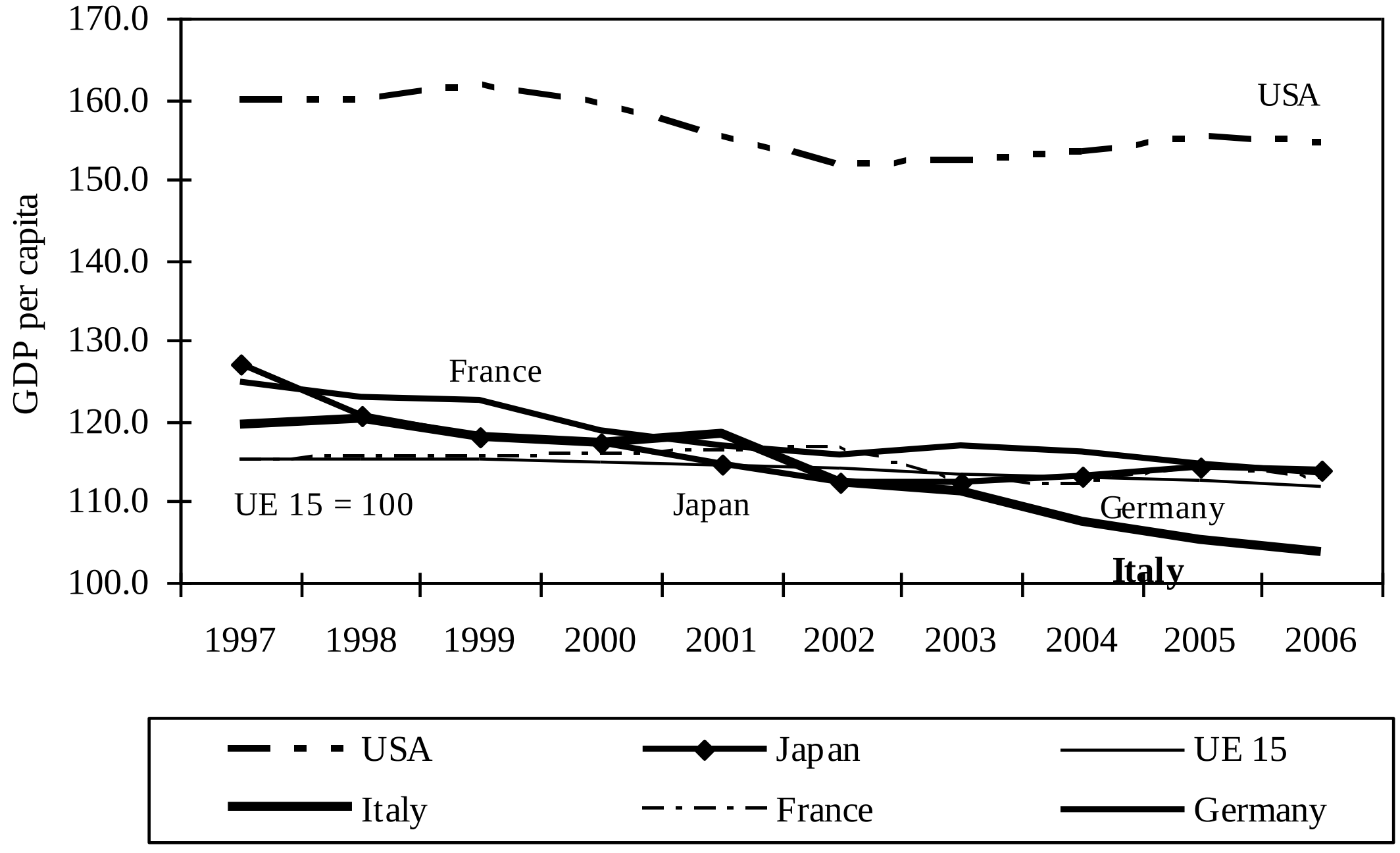

Figure 4 – Trend of GDP per capita over time per country. *Source*: Coccia, 2010b; 2010e

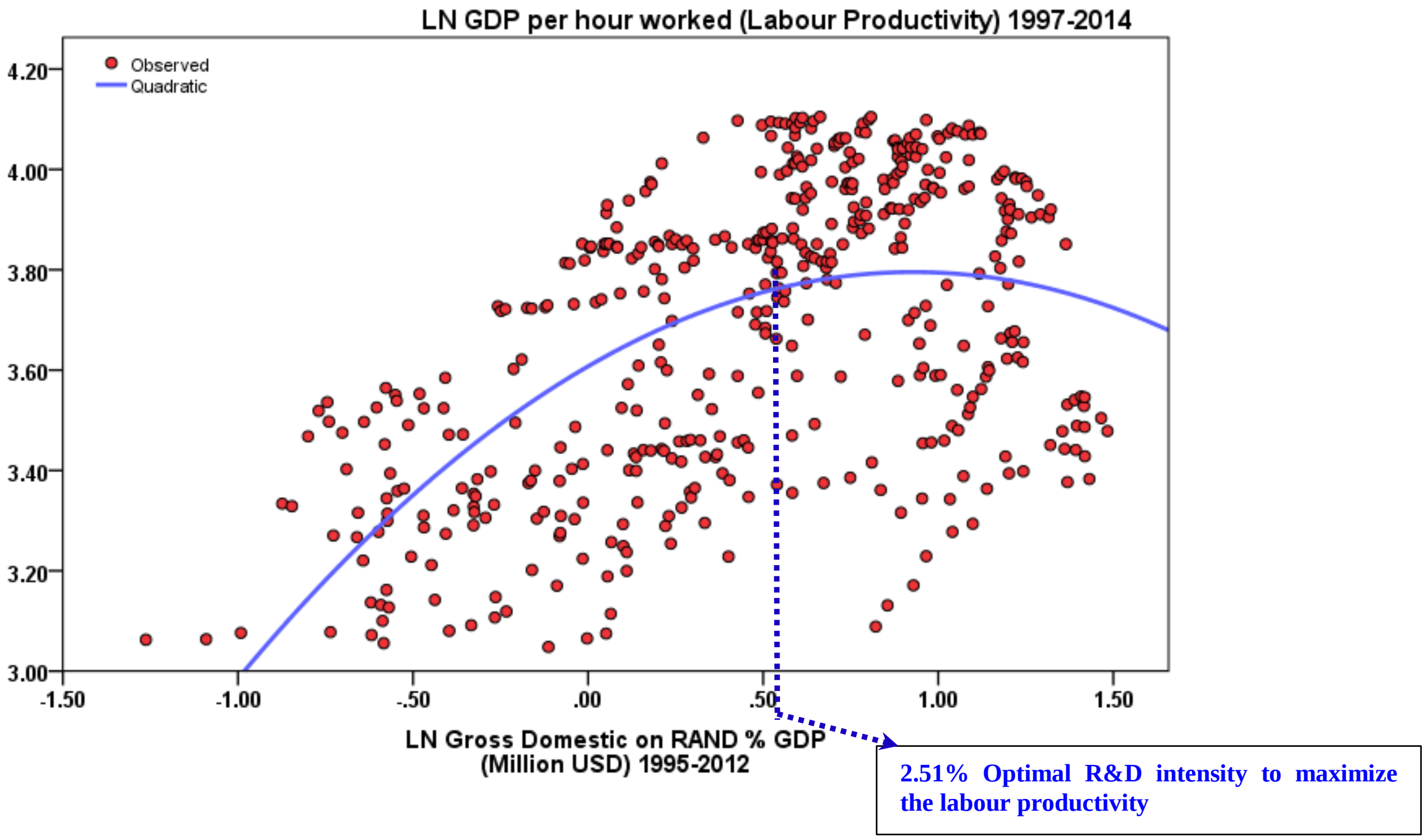

Figure 5 - Curvilinear estimated relationship of LN Labor productivity on LN R&D Investment as percentage of GDP and optimal level of R&D intensity to maximize the labor productivity. *Source*: Coccia M. 2018f. Optimization in R&D intensity and tax on corporate profits for supporting labor productivity of nations, The Journal of Technology Transfer, vol. 43, n. 3, pp. 792-814.

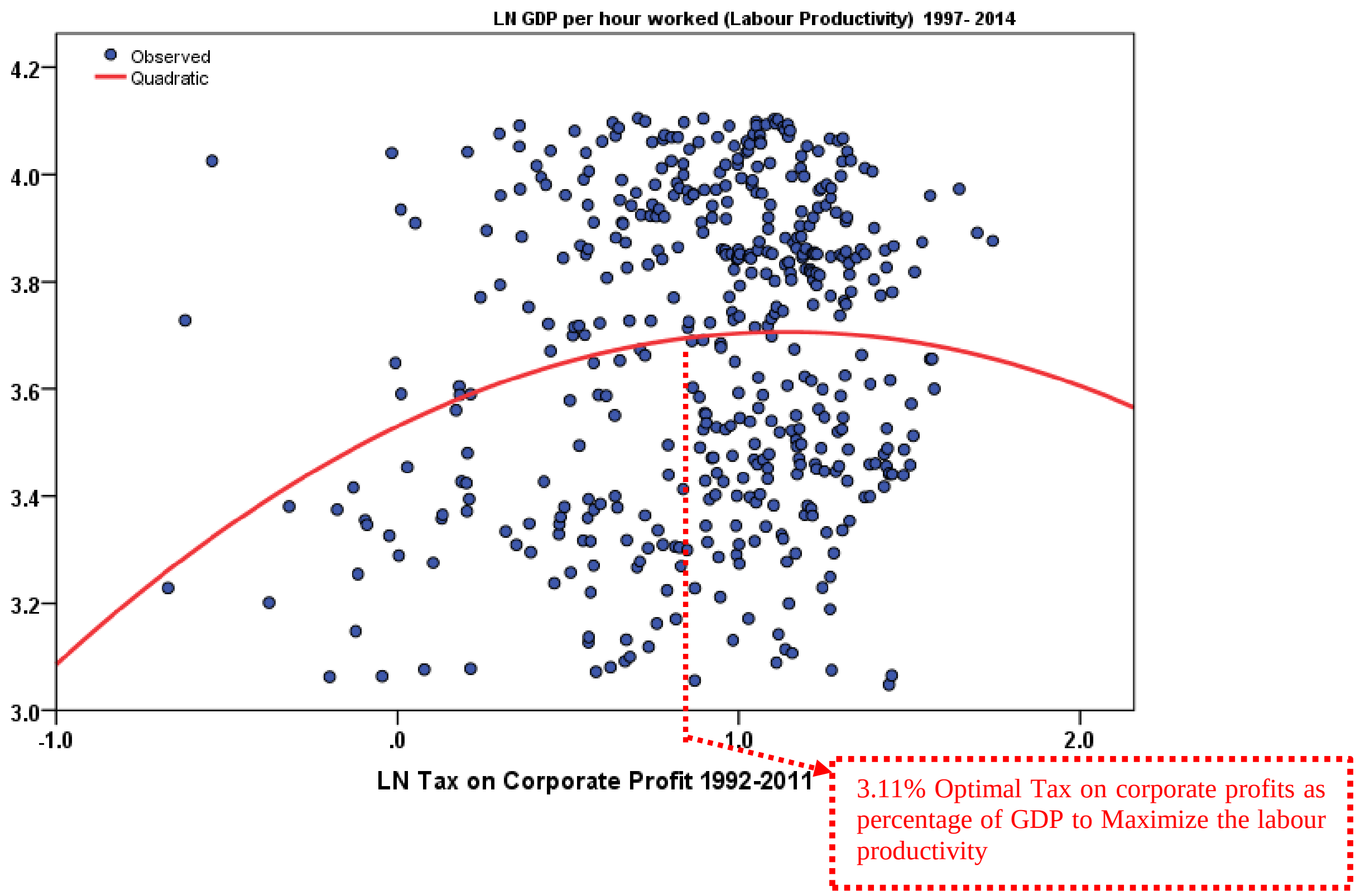

Figure 6- Curvilinear estimated relationship of LN Labour productivity on Tax on corporate profits as percentage of GDP and optimal level of Tax on corporate profits to maximize the labor productivity. *Source*: Coccia M. 2018f. Optimization in R&D intensity and tax on corporate profits for supporting labor productivity of nations, The Journal of Technology Transfer, vol. 43, n. 3, pp. 792-814

Finally, table 3 suggests that leading geoeconomic regions with higher investments in R&D, in particular with higher private R&D expenditure, they foster a higher index of labor productivity.

*Table 3.* Research expenditure (a proxy of investment in science and scientific research) and labor productivity between worldwide players

| World Players | Public R&D Expenditure 1998-2008 * a) | Private R&D Expenditure 1998-2008 * b) | Labor productivity Index 2000=100 (1995-2009) ** |
|---|---|---|---|
| EU (15 countries) | 0.66 (35%)1) | 1.25 (65%) | 101.64 |
| United States | 0.64 (24%) | 1.99 (76%) | 104.88 |
| Japan | 0.73 (23%) | 2.46 (77%) | 103.89 |

*Source:* * Eurostat (2010); ** OECD (2010); *Note*: a) R&D expenditures by government and higher education sector; b) R&D expenditures by business enterprise and private non-profit sector. 1) Percent value of the total.

## Discussion and concluding observations

Bernal (1939) argued that science is considered an “institution” in relation to social and economic events. Bush (1945) claimed that scientific progress is essential to nations and suggested basic principles for governments to support scientific research and higher education. On the basis of the study presented here, the scientific research is a main factor for nations to support socioeconomic power, wealth, economic growth, innovative outputs, etc. Coccia (2018) argues that high investment in scientific research in period of environmental threats can generate general purpose technologies and support long-run economic growth. This study also suggests that nations have a strong incentive to invest in scientific research because long-run consequences are a higher labor productivity and economic growth (cf., Coccia, 2017a).

Overall, then, humankind realized that science and scientific research mean socioeconomic power that in the long run generates many benefits in society (Coccia and Bellitto, 2018). This search for knowledge and investigation of the unknown then became the controlling mechanisms for many research projects in human society. Callon (1994) argues that public investment in R&D is needed to investigate emerging research fields, though results can be uncertain and/or achieved only in the long run, such as studies for measuring gravitational waves and detecting their sources in the universe. In fact, National Science Foundation in the USA has done a huge investment of more than $1 billion for Laser Interferometer Gravitational-Wave Observatory (in construction, operational costs and research funds for scientists) for studying gravitational waves, an unknown research field. In general, the impetus of nations to perform scientific research is to support progress with transfer to techno-economic processes and progressive social change directed to the adaptation of life to new economic and social bases. The interwoven relation between scientific research and new technology yields a greater satisfaction of human needs for improving wellbeing in society. In fact, scientific research of nations supports economic, technological and social change directed to satisfy human wants and human control of nature. Scientific research, combined with technology should be the forerunners of a full realization of the

meaning and possibilities of life of individuals in society (cf., Woods, 1907; Coccia and Bellitto, 2018). Hence, it would be naive to limit the driver of scientific research or at least to make it dependent on the economic vector of nations alone. The scientific research is due to the expanding content of the human life-interests whose increasing realization constitutes progress, rather than external processes conceived in terms of economic processes. Scientific research is a means to support human progress in terms of long-run ideals to satisfy human interests that change in society and characterize the human nature from millennia (Woods, 1907, pp. 813-815; Coccia and Bellitto, 2018). To put it differently, the whole process of scientific research, as reflection of society, is driven by the increasingly effective struggle of the human mind in its efforts to raise superior to the exigencies of the external world, as well as to satisfy human desires, solve problems and achieve/sustain power in society.

To conclude, scientific research is driven by complex factors mainly linked to the question of what human beings truly need and how they seek to address and satisfy real needs and ideals in their social context. This paper shows some determinants of scientific research of nations, such as the goal of achieving socioeconomic power, technological and scientific superiority, higher labor productivity, etc. However, the results and arguments of this study are of course tentative. In fact, the phenomenon is complex and analyses here are not sufficient to understand the comprehensive reasons for and the general implications of science in society, since we know that other things are often not equal over time and space. This preliminary analysis of the reasons inducing nations to perform scientific research may form a ground work for development of more sophisticated studies and theoretical frameworks, focusing on characteristics often neglected in social studies of science. Future efforts in this research field should provide more statistical evidence to support the theoretical framework here. To reiterate, the study here is exploratory in nature and findings need to be considered in light of their limitations. Overall, then, there is need for much more detailed research to shed further theoretical and empirical light on vital determinants supporting scientific research of nations in specific social and contestable environments.